 \documentclass[final,5p,times,twocolumn,authoryear]{elsarticle}

\usepackage{amssymb}
\usepackage{amsmath}
\usepackage{url}

\journal{High Energy Astrophysics}

\begin{document}

\begin{frontmatter}



\title{Cyclotron Line Variability and Accretion Dynamics in Vela X-1}


\author[Tobrej et al.]{
Mohammed Tobrej,$^{1}$\thanks{tabrez.md565@gmail.com}
Binay Rai,$^{1}$\thanks{binayrai21@gmail.com}
Manoj Ghising,$^{1}$\thanks{manojghising26@gmail.com}
Bikash Chandra Paul$^{1}$\thanks{bcpaul@associates.iucaa.in}
\\
$^{1}$Department of Physics, North Bengal University, Siliguri, Darjeeling, WB, 734013, India
\\
}

\begin{abstract}
We present a comprehensive analysis of Vela X-1 using two new \textit{NuSTAR} observations, placed in the context of four earlier datasets obtained between 2012 and 2020.  The energy-resolved pulse profiles demonstrate a significant transformation from an asymmetric structure at low energies to distinct double peaks above $\sim$ 12 keV, whereas the pulse fraction escalates with photon energy but decreases with flux.  Broadband spectra validate the Fe $K_\alpha$ emission line and disclose both fundamental and harmonic cyclotron resonant scattering characteristics (CRSF).  We observe no substantial link between CRSF energies and luminosity, contrary to previous findings; rather, the photon index and folding energy demonstrate distinct anti-correlations with flux, aligning with sub-critical accretion and increased Comptonization in the accretion column.  Our results provide the first clear evidence that the harmonic CRSF in Vela X-1 does not follow the long-term decay previously claimed.  The fundamental line energy also displays an irregular evolution, without a clear monotonic trend. Notably, the harmonic-to-fundamental energy ratio departs from the canonical value of two, suggesting that the line-forming regions are located at different heights within the accretion column. These results provide new constraints on the accretion geometry and magnetic field topology of Vela X-1, highlighting the importance of continued monitoring with current and future X-ray observatories.
\end{abstract}

%




\end{frontmatter}




\section{Introduction}
Vela X-1 is a persistent high-mass X-ray binary (HMXB) consisting of a neutron star and the B0.5\,Ib supergiant HD~77523 \citep{Hiltner}. The neutron star, with a spin period of $\sim$283\,s, orbits its companion every 8.9 days in a mildly eccentric orbit ($e \sim 0.09$). Due to its proximity ($d = 1.99 \pm 0.13$\,kpc; \citep{Kretschmar}, it remains one of the brightest wind-fed accreting X-ray pulsars despite its moderate intrinsic luminosity \citep{Nagase,Furst}. With a dynamically measured mass exceeding 1.8\,$M_\odot$ \citep{van Kerkwijk,Quaintrell}, the system also provides important constraints on the neutron-star mass distribution and the equation of state of dense matter.

The stellar wind from the donor is known to be clumpy \citep{Owocki,Dessart,Ducci,Furst,Oskinova}, producing strong variability in absorption and accretion rates \citep{Haberl, Oskinova}. Typical clump properties inferred from \textit{INTEGRAL}, \textit{XMM-Newton}, and \textit{Suzaku} observations include masses of $\sim 5\times10^{19}$\,g and radii of $\sim 2\times10^{10}$\,cm \citep{Furst, Odaka, Martinez}. The resulting X-ray emission exhibits a complex continuum spectrum shaped by Comptonization in the accretion column, together with cyclotron resonant scattering features (CRSFs) that directly probe the magnetic field strength \citep{Meszaros}. The centroid energy of the fundamental line is related to the local magnetic field by
\begin{equation}
E_{\rm cyc} \simeq 11.6\,B_{12}\,(1+z)^{-1} \ \mathrm{keV},
\end{equation}
where $B_{12}$ is the field in units of $10^{12}$, G and $z\sim0.3$ is the gravitational redshift for the parameters of the canonical neutron star. 

 Vela X-1 has been extensively studied with \textit{RXTE}, \textit{Suzaku}, and \textit{NuSTAR}, which revealed both the fundamental and harmonic CRSFs at \textbf{$\sim$23--25\,keV and $\sim$54--57\,keV}, respectively \citep{Kreykenbohm, La Barbera, furstn, Diez}. Previous works suggested a possible long-term decrease in the harmonic CRSF energy \citep{La Parola, Ji}, although this interpretation remains debated. Moreover, the harmonic-to-fundamental ratio in Vela X-1 often departs from the canonical value of two, raising questions about the geometry of the line-forming regions \citep{Nishimura, Schonherr}. More recent studies have further explored the properties of Vela~X-1, including polarimetric measurements with \textit{IXPE} \citep{Forsblom} and high-resolution spectroscopy from the first \textit{XRISM} observations \citep{diez2025}.

This work is based on the timing and spectral analysis of two newly examined \textit{NuSTAR} observations of Vela X-1 from 2020, contextualized using four prior observations acquired between 2012 and 2020. Our goals are to (i) investigate the energy dependence of pulse profiles and pulse fractions, (ii) examine the evolution of CRSF parameters with flux and time, and (iii) test the long-term behavior of both the fundamental and harmonic lines. Particular attention is paid to the 2020 observations, which provide new insights into the anharmonic spacing of CRSFs and the absence of a persistent secular trend. These results allow us to place new constraints on the accretion dynamics and magnetic field structure of Vela X-1.

\section{Observation and Data reduction}
The \textit{Nuclear Spectroscopic Telescope Array (NuSTAR)} operates in the broad energy range of 3--79 keV and is the first focusing hard X-ray telescope in orbit. It comprises two co-aligned grazing-incidence telescopes that are similar but not identical, each associated with a focal plane module, FPMA and FPMB, housing pixelated CdZnTe solid-state detectors \citep{15}. With its unprecedented sensitivity and high spectral resolution in the hard X-ray band, \textit{NuSTAR} is particularly well suited for investigating the characteristic absorption features in Vela X-1.  

The clean event files used for our analysis were generated with the mission-specific task \texttt{NUPIPELINE}. The resulting event files were examined using the \texttt{XSELECT} tool, while imaging and visualization were performed with the DS9 package\footnote{\url{https://sites.google.com/cfa.harvard.edu/saoimageds9}}. Source spectra and light curves were extracted from a circular region of radius $150^{\prime \prime}$ centered on detector 1, while background events were accumulated from an $80^{\prime \prime}$ circular region on detector 2, located away from the source. The task \texttt{NUPRODUCTS} was employed to generate the final source and background spectra, response matrices, and light curves. Background-subtracted light curves were produced using the \texttt{LCMATH} tool, and barycentric corrections were applied to the photon arrival times using the \texttt{BARYCORR} task.  

\begin{figure}

\begin{center}
\includegraphics[angle=0,scale=0.25]{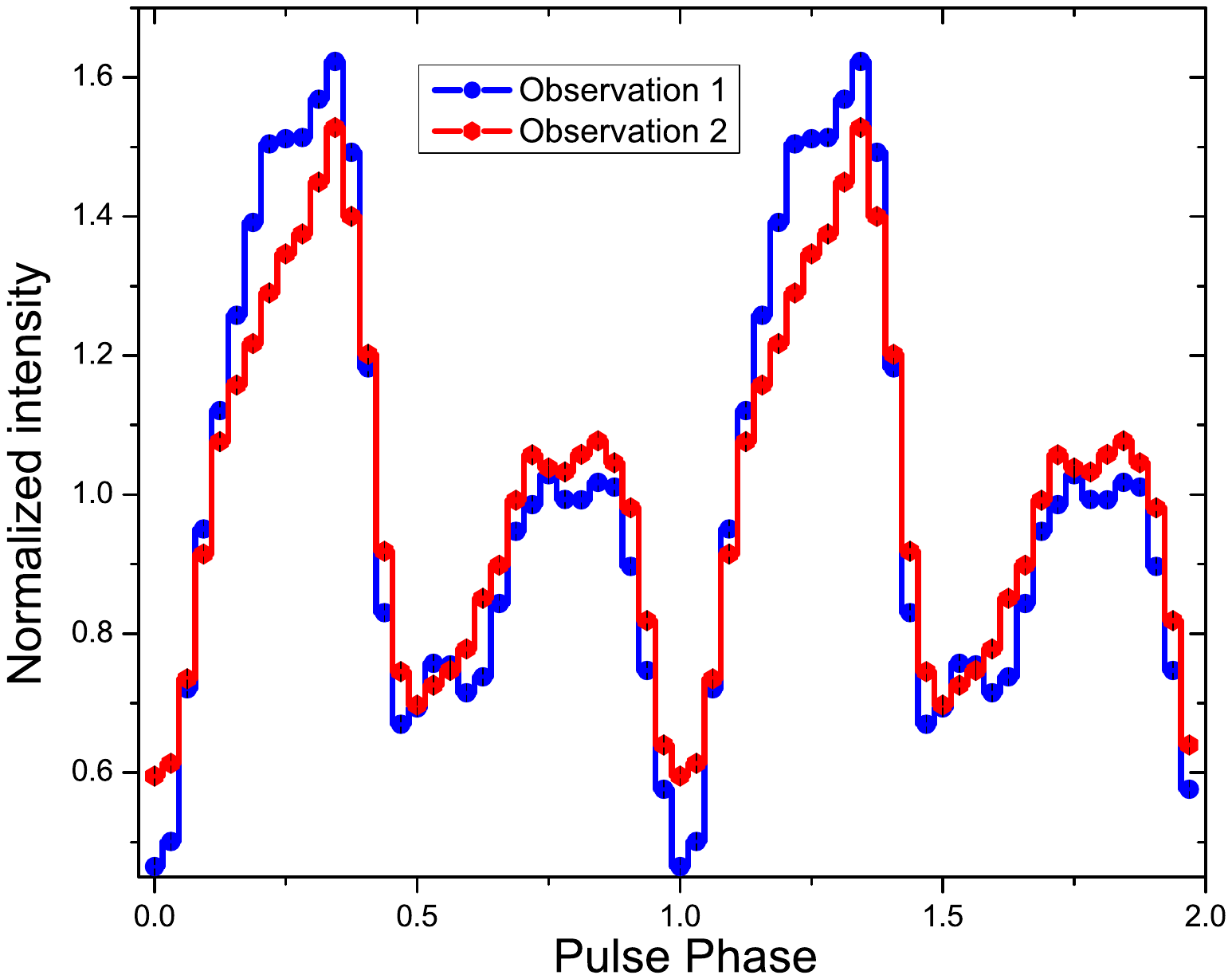}
\end{center}

\caption{Pulse profile of Vela X-1 in the 3--79\,keV band obtained from \textit{NuSTAR} data, folded at the best-fit pulse period using 32 phase bins. Profiles are normalized to the mean count rate.}
\label{1}
\end{figure}

We analyzed two NuSTAR observations of Vela X-1 from 2020, with ObsIDs 90602328004 and 90602328006 , processed using standard procedures. Spectral fitting was carried out with \texttt{XSPEC} version 12.12.1 \citep{s}. The orbital phases corresponding to the \textit{NuSTAR} observations were determined using the ephemeris of \citet{Kreykenbohm} and are listed in Table\ref{1}. 

\begin{table*}
\begin{center}
\begin{tabular}{ccccccc}
\hline
Observation& Observation ID &  Observation Date  &	Exposure (ks) & Orbital Phase & Reference\\
 &  &(DD-MM-YYYY)   & &covered \\	
I&90602328004 &	2020-09-26 &		9.59 & (0.49-0.52)  &This work  \\
II&90602328006 &	2020-09-29 &		8.98 & (0.79-0.82)&This Work    \\
\hline
III&90402339002 &   2019-01-10 &        36.04 & (0.68-0.78)& \cite{Diez}\\
IV&30501003002 &   2019-01-10 &        40.38 & (0.36–0.52)& \cite{Diez}\\
V&30002007003 &   2013-04-22 &        41.67 & (0.66–0.77)& \cite{furstn}\\
VI&10002007001 &   2012-07-09 &        10.81 & (0.66-0.69)&\cite{furstn}\\

\hline
\end{tabular}

\caption{NuSTAR observations represented by the observation IDs along with the date of observation, exposure, and orbital phases.}  
\end{center}
\end{table*}

\section{Timing Analysis}
\begin{figure*}
\begin{minipage}{0.3\textwidth}
\includegraphics[angle=0,scale=0.29]{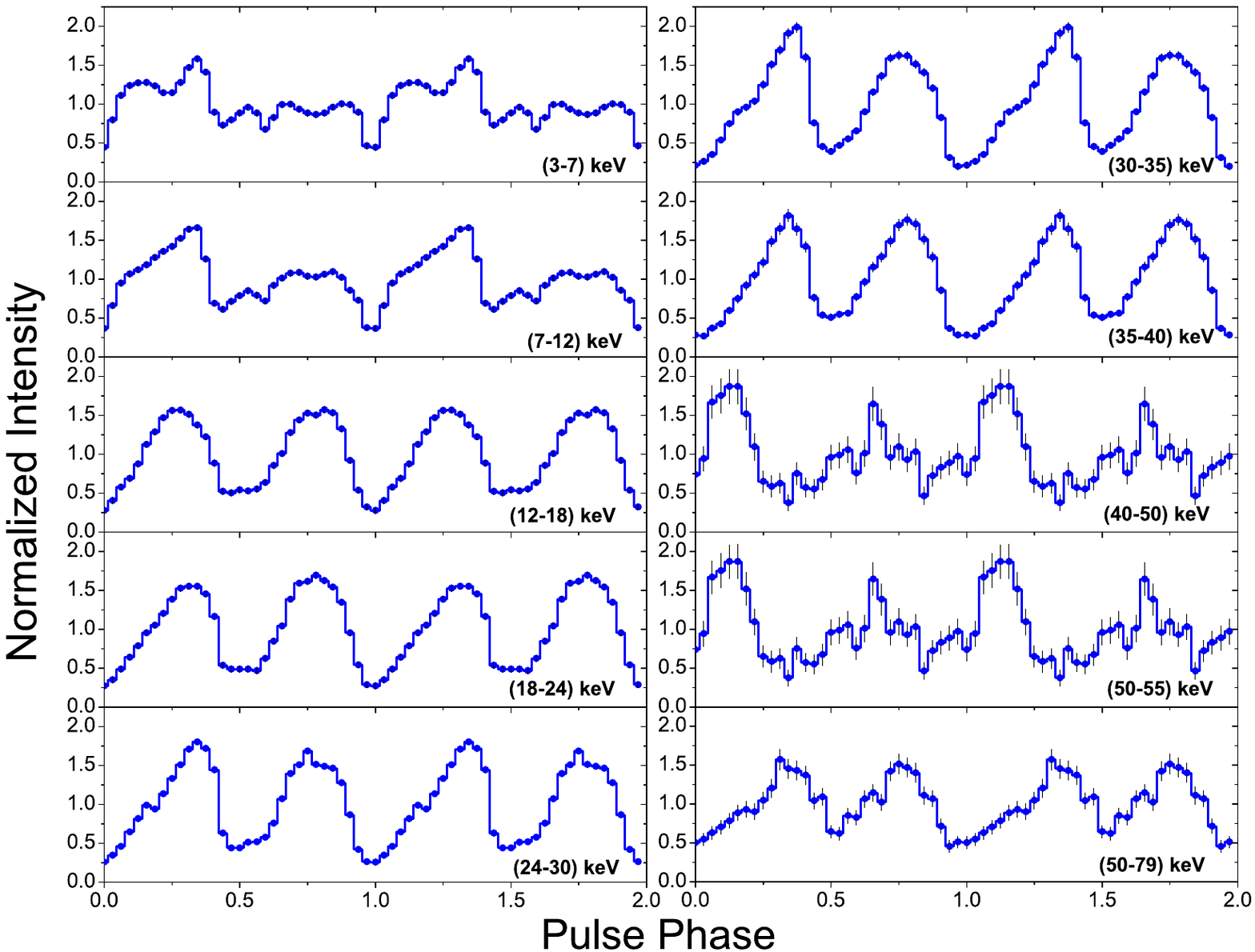}
\end{minipage}
\hspace{0.2\linewidth}
\begin{minipage}{0.3\textwidth}
\includegraphics[angle=0,scale=0.29]{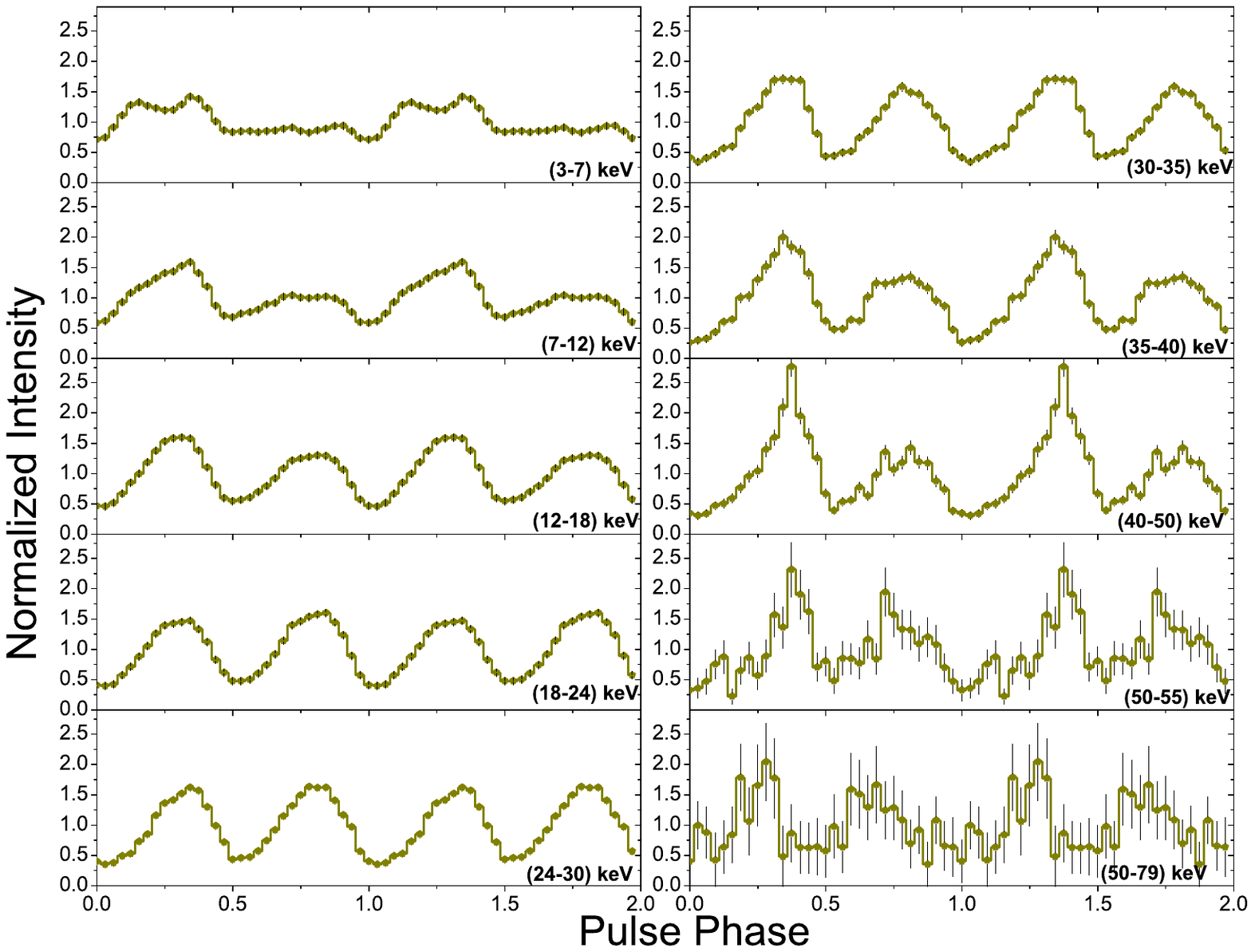}
\end{minipage}
\caption{Energy-resolved pulse profiles of Vela X-1 for Observation~I (left) and Observation~II (right). Each panel corresponds to a different energy band, illustrating the transition from asymmetric shapes at low energies to double-peaked structures above $\sim$12\,keV.}
\label{2}
\end{figure*}

\begin{figure}

\begin{minipage}{0.3\textwidth}
\includegraphics[angle=0,scale=0.25]{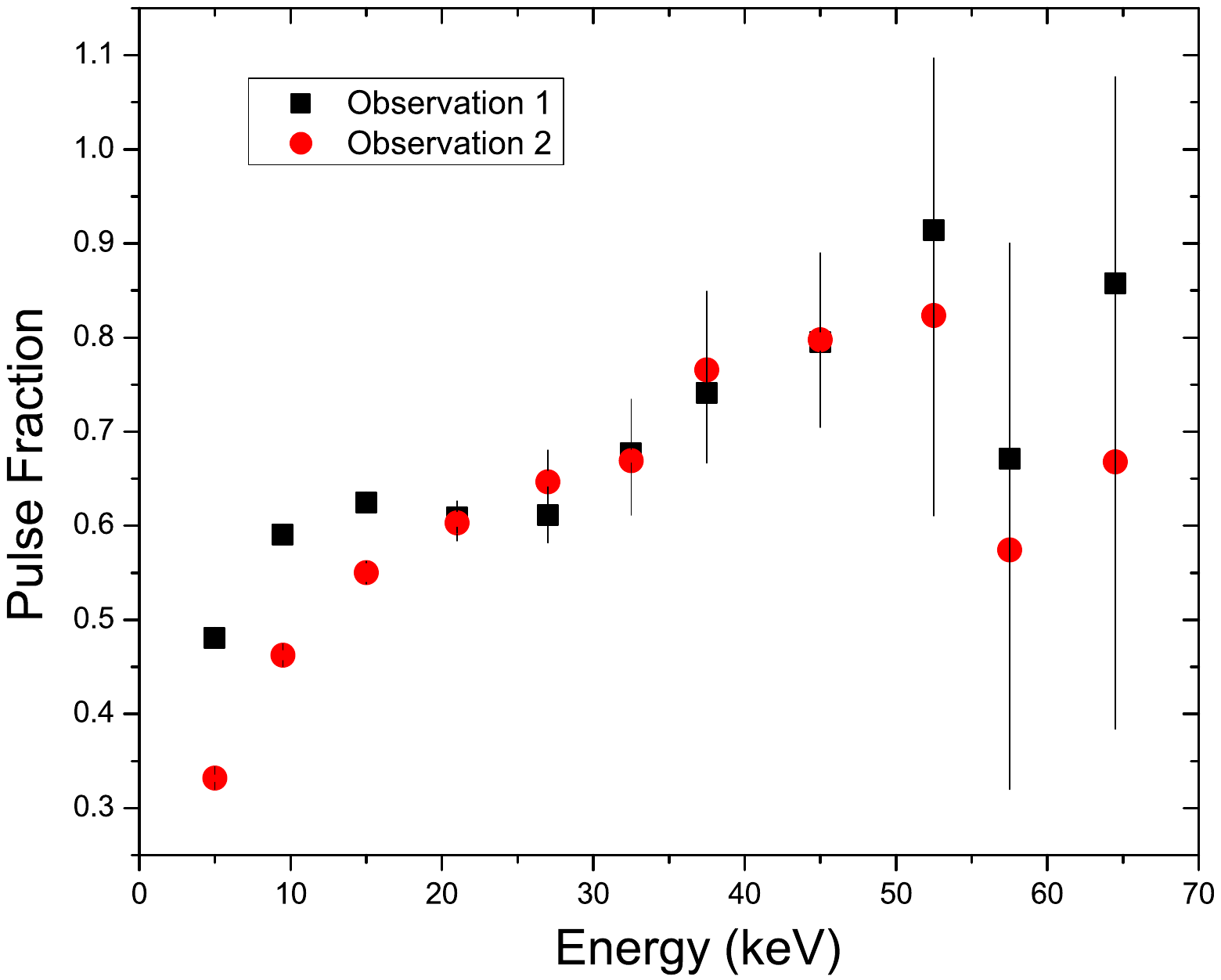}
\end{minipage}
\hspace{0.2\linewidth}
\begin{minipage}{0.3\textwidth}
\includegraphics[angle=0,scale=0.32]{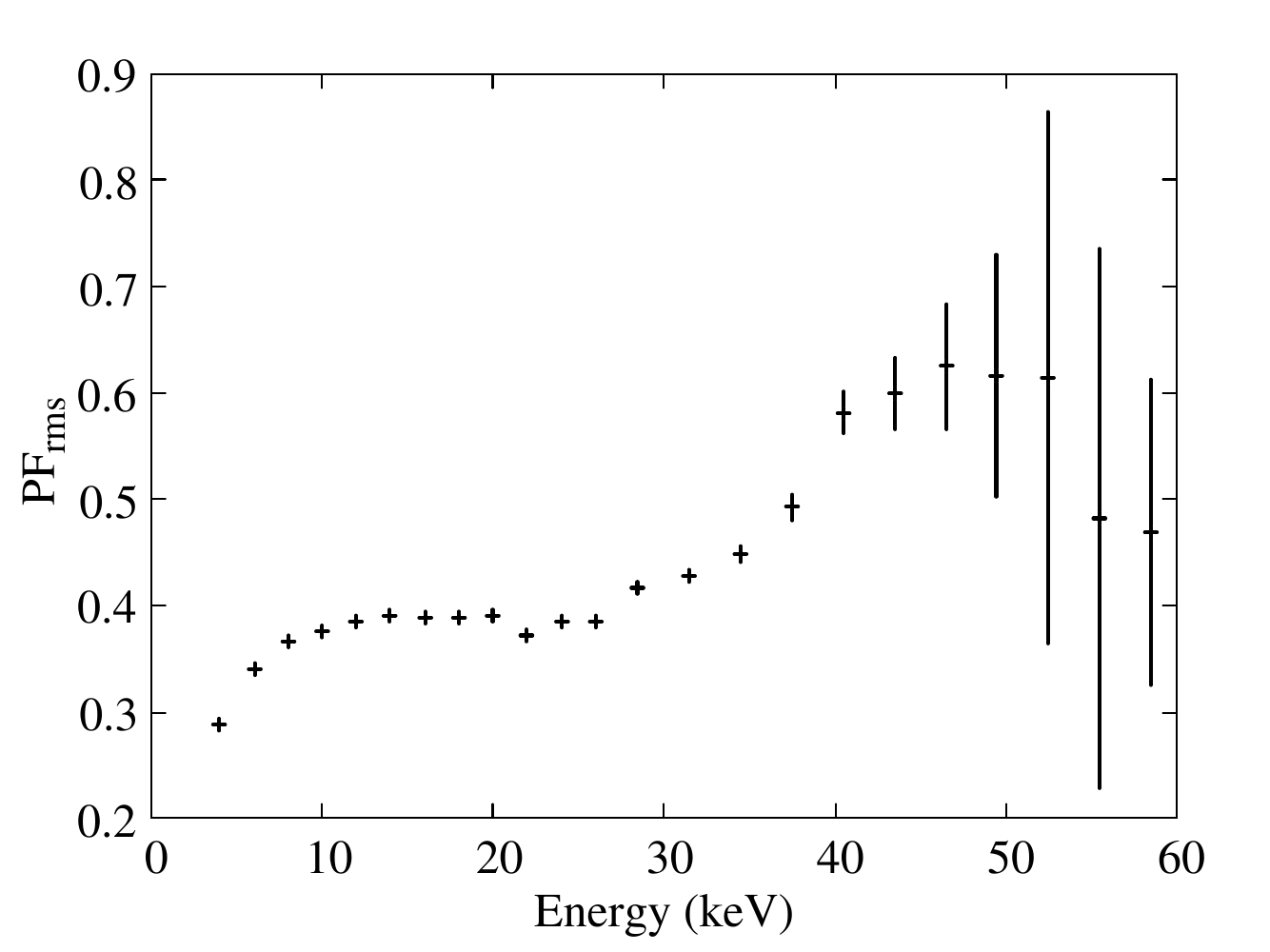}
\end{minipage}

\caption{Pulse fraction versus photon energy for Observations I and II (top), and RMS Pulse fraction versus energy for Observation I (bottom). The pulse fraction increases with energy, as typically seen in accreting pulsars, with noticeable deviations around the CRSF energy.}
\label{3}
\end{figure}
\begin{figure}

\begin{center}
\includegraphics[angle=0,scale=0.25]{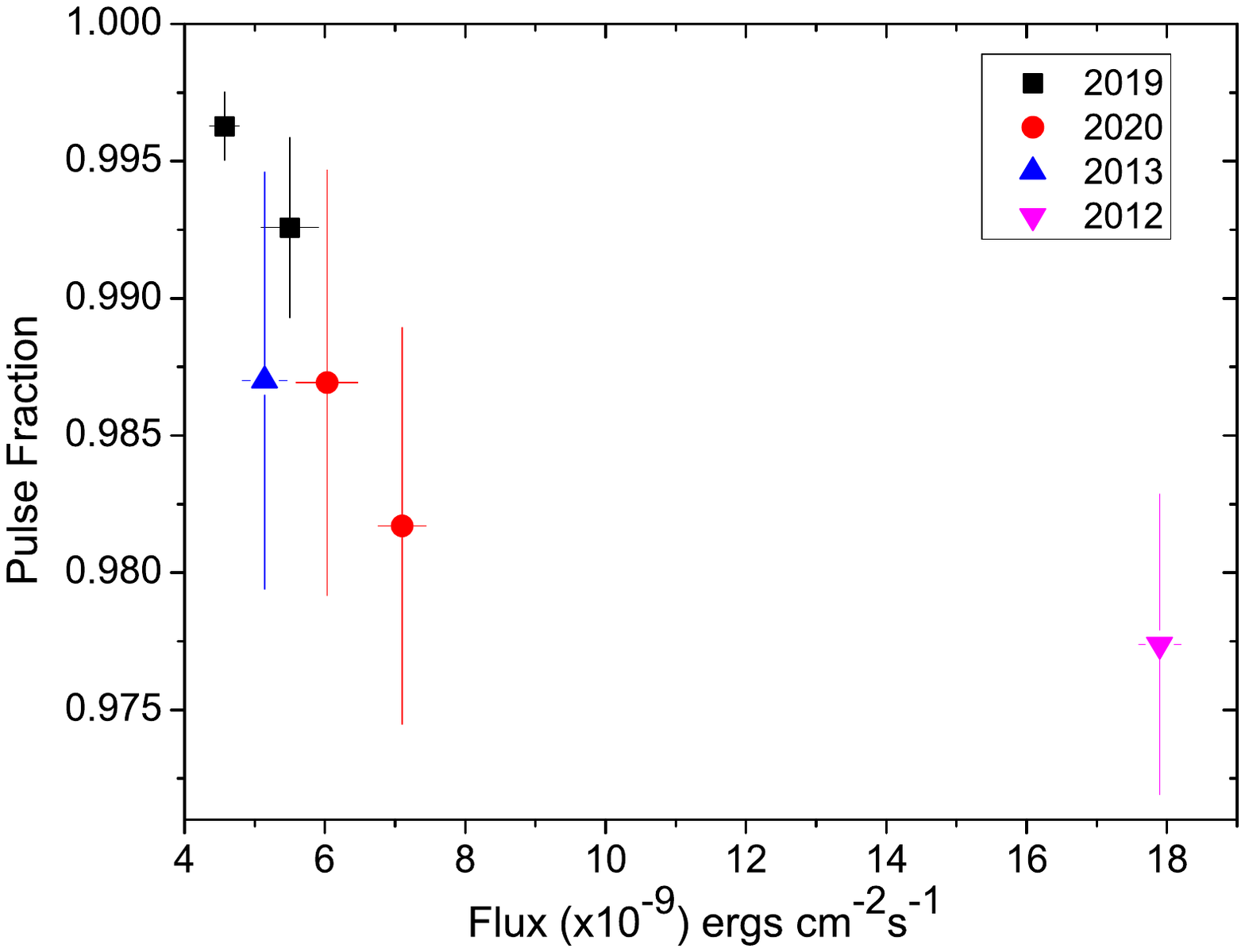}
\end{center}
\caption{Pulse fraction versus flux for all six \textit{NuSTAR} observations (2012--2020) in 3–79 keV range. A moderate decreasing trend is evident (r=-0.6), consistent with enhanced unpulsed emission at higher accretion rates.}
\label{4}
\end{figure}

\subsection{Timing Analysis}

For the temporal analysis, \textit{NuSTAR} light curves were extracted with a bin size of 0.5 s using the FTOOL \texttt{LCURVE}. A preliminary estimate of the source pulsation was obtained by applying a Fast Fourier Transform (FFT) to the light curves. To refine the measurement, the epoch-folding technique \citep{16,17}, based on maximization of $\chi^{2}$, was used. The best-fit pulse periods were determined to be $283.48 \pm 0.13$ s and $283.43 \pm 0.13$ s for the two observations, respectively.  The uncertainties \textbf{(1-$\sigma$)} in the pulse period measurements were estimated following the method of \citet{Boldin}. We generated 1000 simulated light curves, measured the corresponding pulse periods, and calculated the standard deviation and standard error to quantify the uncertainty. The FTOOL \texttt{EFSEARCH} was used to determine the best pulse period, while the folded pulse profiles were subsequently obtained with the tool \texttt{EFOLD}.

\subsection{Light curves, Pulse profiles and Pulse fraction}
The broad energy coverage of \textit{NuSTAR} (3--79 keV) has been utilized to investigate the X-ray variability of Vela X-1. To study the emission geometry at different energies, we generated energy-resolved pulse profiles. The profiles show a strong dependence on both energy and time: at lower energies they appear asymmetric, whereas at higher energies they evolve into prominent double-peaked structures. With increasing energy, the secondary peak gradually grows in amplitude, while the relative strength of the primary peak also increases.  

The light curves were divided into energy bands of 3--7 keV, 7--12 keV, 12--18 keV, 18--24 keV, 24--30 keV, 30--35 keV, 35--40 keV, 40--50 keV, 50--55 keV, and 55--79 keV. The light-curve energy bands were selected to balance spectral coverage with adequate signal-to-noise across the \textit{NuSTAR} range.  The 50--55\,keV interval was included to sample the region immediately below the known $\sim55$\,keV cyclotron harmonic, avoiding dilution within the wider 55--79\,keV band. Although not based on a specific physical model, this  provides sufficient S/N while retaining finer sampling near the harmonic. Folding was performed using a reference epoch $T_{0}$, chosen such that 
the minimum-flux bin of the full $3$--$79~\mathrm{keV}$ light curve was 
aligned at phase zero. Pulse profiles in all sub–energy bands were then 
shifted using this same $T_{0}$ to ensure a consistent phase reference 
across energies. Above 12 keV, the initially distorted profiles transition into well-defined double-peaked patterns. In particular, within the 18--24 keV band, the secondary peak becomes stronger than the primary peak in observation I, while the two peaks are nearly equal in observation II. Variations in the count rate across different bands make a direct comparison of modulation strengths challenging. The overall evolution of the profiles indicates changes in the accretion geometry of the pulsar. The folded profile in the 3--79 keV band and the energy-resolved profiles are shown in Figure \ref{1} and Figure \ref{2}, respectively.  

The pulse fraction (PF), which quantifies the modulation amplitude of the profiles, is defined as  
\begin{equation}
PF = \frac{P_{\rm max} - P_{\rm min}}{P_{\rm max} + P_{\rm min}},
\end{equation}  
where $P_{\rm max}$ and $P_{\rm min}$ are the maximum and minimum intensities of the profile, respectively \citep{w}. PF provides insight into the relative contributions of pulsed emission from the accretion column and unpulsed emission from other regions such as the neutron star surface.  

For Vela X-1, the PF increases with energy, a behavior commonly observed in accreting X-ray pulsars \citep{37}. At lower energies, the PF of observation I is significantly higher than that of observation II, while above 20 keV the two observations yield comparable values. However, at energies $>50$ keV, the PF of observation I again exceeds that of observation II. The increasing PF with energy can be explained by the toy model of \citet{landt}, where the X-ray emitting region contracts with energy, leading to enhanced pulsed emission. In addition, localized decreases in PF near specific energies are consistent with the presence of absorption features in the continuum. In particular, a reduction in PF around the cyclotron line energy has been reported in several X-ray pulsars \citep{TSY,39,37,tsy,lut}. The evolution of PF with energy is shown in Figure \ref{3}. \cite{Ferrigno} introduced a method to extract spectral information from energy-resolved pulse profiles by modeling the energy dependence of the pulsed fraction, allowing features such as iron lines or cyclotron absorption to be probed via PF spectroscopy. The RMS-based approach is particularly effective in tracing the pulsed fraction, as it is less affected by low-count statistics, making it well suited to identify energy-dependent features seen in spectral analysis. The RMS pulsed fraction, $PF_{\mathrm{rms}}$, is computed directly from the root-mean-square of the pulse-profile bins as :

\begin{equation}
\mathrm{PF}_{\mathrm{rms}}
= \frac{1}{\bar{p}}
\sqrt{
\frac{1}{N}
\sum_{i=1}^{N}
\left(p_i - \bar{p}\right)^2
}
\end{equation}
where $\bar{p}$ is the mean count rate, $p_i$ is the count rate in the $i$-th phase bin.

As evident from figure\ref{3}, the  RMS pulsed fraction  exhibits a non-monotonic energy dependence, likely reflecting changes in emission geometry and beaming with energy. By employing comparatively narrower energy bands, the results remain consistent and clearly show deviations around the cyclotron line energies, which  account for the observed drop in the pulsed fraction.

We further computed PF values across all available \textit{NuSTAR} observations in the 3--79 keV range and examined their variation with flux. As illustrated in Figure \ref{4}, the PF exhibits an overall decreasing trend with increasing flux (or luminosity).

\section{Spectral Analysis}

\begin{figure*}
\begin{center}
\includegraphics[angle=-90, scale=0.6]{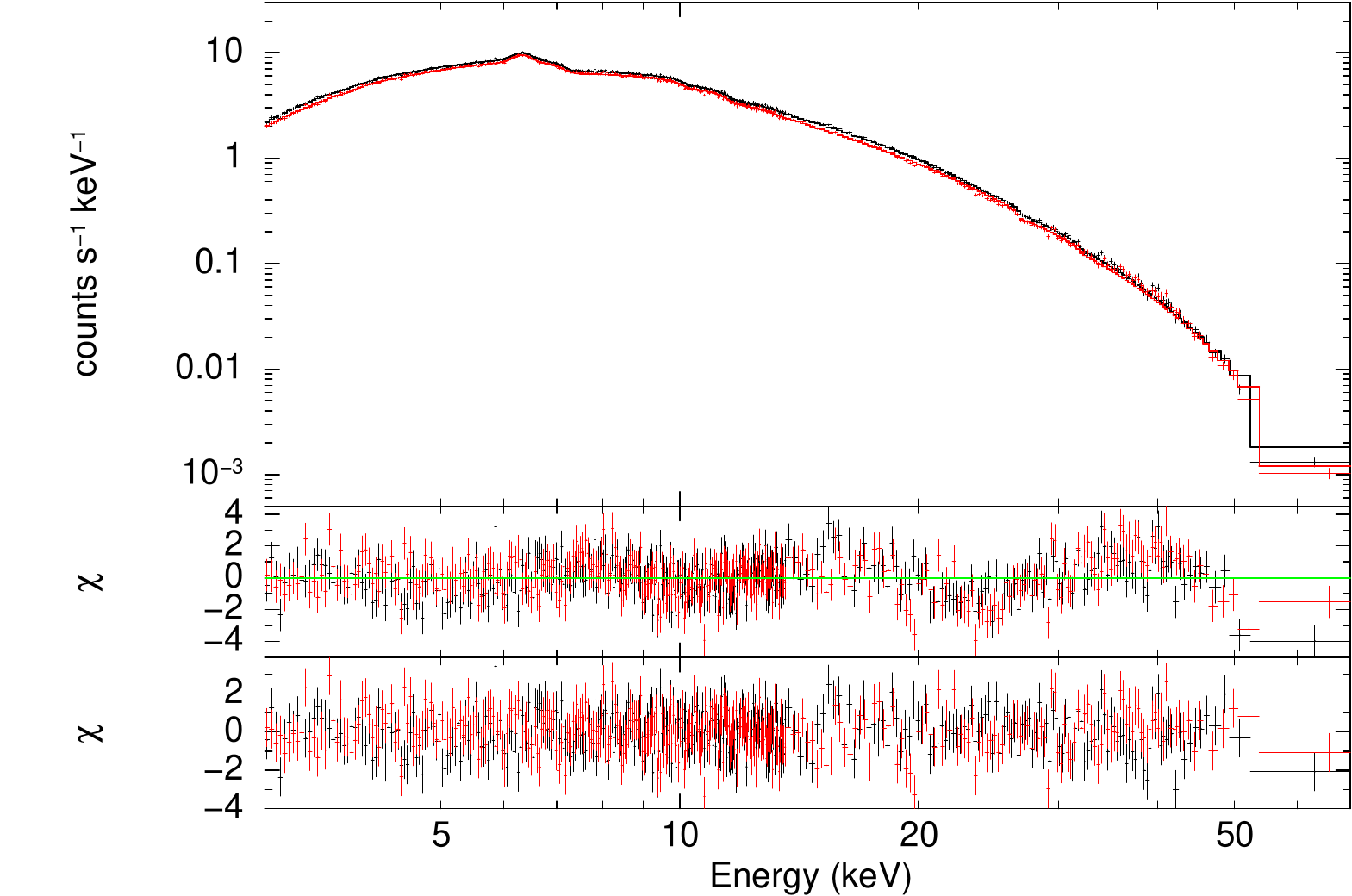}
\end{center}
\caption{\textit{NuSTAR} spectra of Vela X-1 (FPMA: black, FPMB: red) in the 3--79\,keV range. Top: unfolded spectra. Middle: residuals without CRSF components. Bottom: residuals after including Gaussian absorption lines (GABS). Spectra are rebinned for representation.}
\label{5}
\end{figure*}

The \textit{NuSTAR} spectra of Vela X-1 from both FPMA and FPMB were fitted in the broad 3--79 keV energy range. The spectra are shown in Figure \ref{5}. We performed the spectral fitting using the optimal binning approach proposed by \cite{Kaastra}. To account for relative normalization between the modules, the constant factor for FPMA was fixed at unity, while that for FPMB was left free. The resulting cross-normalization constants were $1.01 \pm 0.002$ and $0.99 \pm 0.003$ for the two observations, consistent with the expected 1--2\% calibration uncertainty \citep{E}. 

For the continuum, we used a power law with photon index $\Gamma$ modified by a Fermi--Dirac cutoff (FDcut; \citealt{Tanaka}), a model that has been extensively employed in the study of accreting X-ray pulsars and CRSF sources \citep[e.g.][]{Makishima, Coburn}. The FDcut form is defined as
\begin{equation}
    F(E) \propto E^{-\Gamma} \left[1 + \exp\left(\frac{E - E_{\rm cut}}{E_{\rm fold}}\right)\right]^{-1},
\end{equation}
where $E_{\rm cut}$ and $E_{\rm fold}$ represent the cutoff and folding energies, respectively.

Photoelectric absorption was modeled with the \texttt{TBABS} component, adopting the abundances of \citet{Wilms2000} and cross-sections from \citet{Verner}. A single absorber did not adequately describe the soft X-ray curvature, likely reflecting additional absorption from the structured, clumpy stellar wind in the system. We therefore included a partial-covering component (\texttt{TBPCF}), a standard approach for wind-fed HMXBs \citep[e.g.][]{furstt, Grinberg}. The interstellar absorption column was fixed to $N_{\rm H} = 0.371 \times 10^{22}$ cm$^{-2}$, consistent with previous studies, while the partial covering absorber was left free. Since \textit{NuSTAR} is not sensitive below 3 keV, the lower $N_{\rm H}$ values could not be constrained.

A positive residual near 6.4 keV was modeled with a Gaussian component, representing Fe K$_\alpha$, with centroid $6.343 \pm 0.062$ keV and equivalent width $0.110 \pm 0.001$ keV. The continuum emission of HMXB pulsars is understood to arise from Comptonization of soft photons in the accretion column plasma above the neutron star surface. To account for the negative residuals, we introduced Gaussian absorption components (\texttt{GABS}). The first significant feature appears at around (23$-$25)\,keV, consistent with the fundamental cyclotron line reported in the literature. Adding a second \texttt{GABS} component at  (54$-$57)\,keV, corresponding to the expected first harmonic, further improved the fit significantly. Residuals also remained around $10$--$11$\,keV. We modeled this feature with an additional absorption component; however, we stress that this structure is \emph{not interpreted as a physical CRSF}. Similar  residuals have been reported in several CRSF systems, including Vela~X-1 \citep{La Barbera, Kreykenbohm, furstt} and other accreting pulsars \citep{Coburn, Tamang}. A thorough investigation of the 10 keV feature has also been presented by \cite{Manikantan}. As noted by \citet{E}, the \textit{NuSTAR} response contains small calibration uncertainties near the tungsten L-edge at $\sim 10$\,keV. \textbf{Since similar 10 keV feature has been observed with other instruments (e.g., Suzaku), a purely instrumental origin is unlikely. No generally accepted physical interpretation currently exists, and the feature may be intrinsic to the source.}

 The final model therefore included three Gaussian absorption components. This produced an excellent fit with $\chi^{2}$/dof = 1173.9/1083 = 1.08. The best-fit photon index was $\Gamma=\;1.16\pm0.05$. The partial covering absorber gave $N_{\rm H} = (17.8 \pm 0.7) \times 10^{22} cm^{-2}$. The unabsorbed flux in the 3--79 keV range was $ (7.1\pm0.3) \times 10^{-9} erg \;cm^{-2}\; s^{-1}$, corresponding to a luminosity of  3.4 $\times 10^{36} erg\; s^{-1}$ at a distance of 1.99 kpc \citep{Kretschmar}.  For observation II, the final continuum model provided $\chi^{2}$/dof = 1157.6/1048 = 1.10. The best-fit parameters for both observations are summarized in Table 2.  

 To determine the significance of the CRSF lines, we simulated $10^{4}$ spectra \textbf{for observation I} using \texttt{simftest} and fitted each simulated spectrum with the continuum model both with and without the CRSF components. From these fits, we obtained the distribution of $\Delta\chi^{2}$ values expected under the null hypothesis. The observed $\Delta\chi^{2}$ value of $614.313$  corresponding to the fundamental and harmonic lines, respectively, are far larger than any $\Delta\chi^{2}$ produced in the simulations. This indicates an extremely small F-test probability and a significance well above conventional detection thresholds. In fact, the probability of obtaining such large $\Delta\chi^{2}$ values under the null hypothesis is $<10^{-4}$, corresponding to a detection significance of $>4\sigma$.

\begin{figure*}
\begin{minipage}{0.30\textwidth}
\includegraphics[angle=0,scale=0.2]
{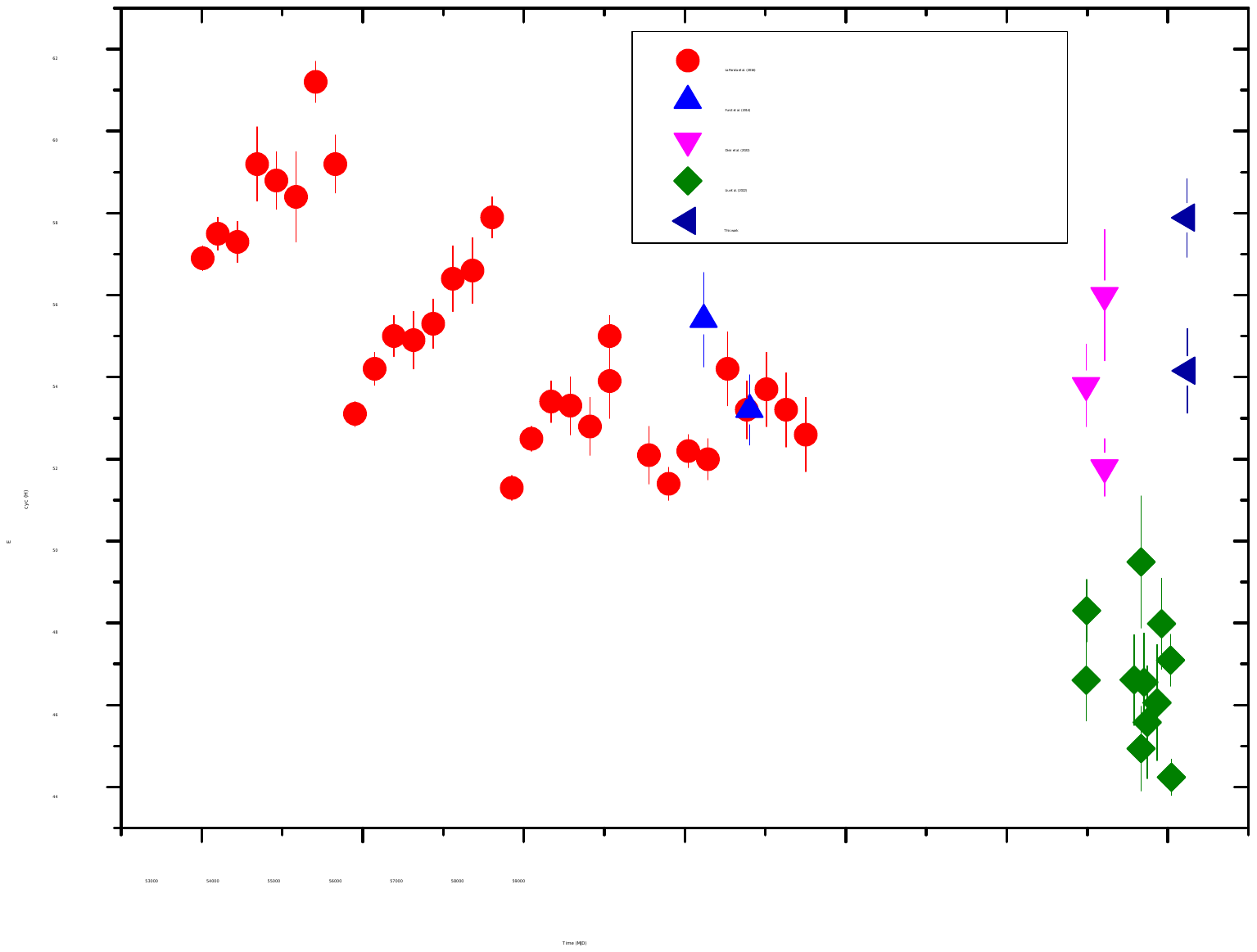}
\end{minipage}
\hspace{0.04\linewidth}
\begin{minipage}{0.30\textwidth}
\includegraphics[angle=0,scale=0.25]{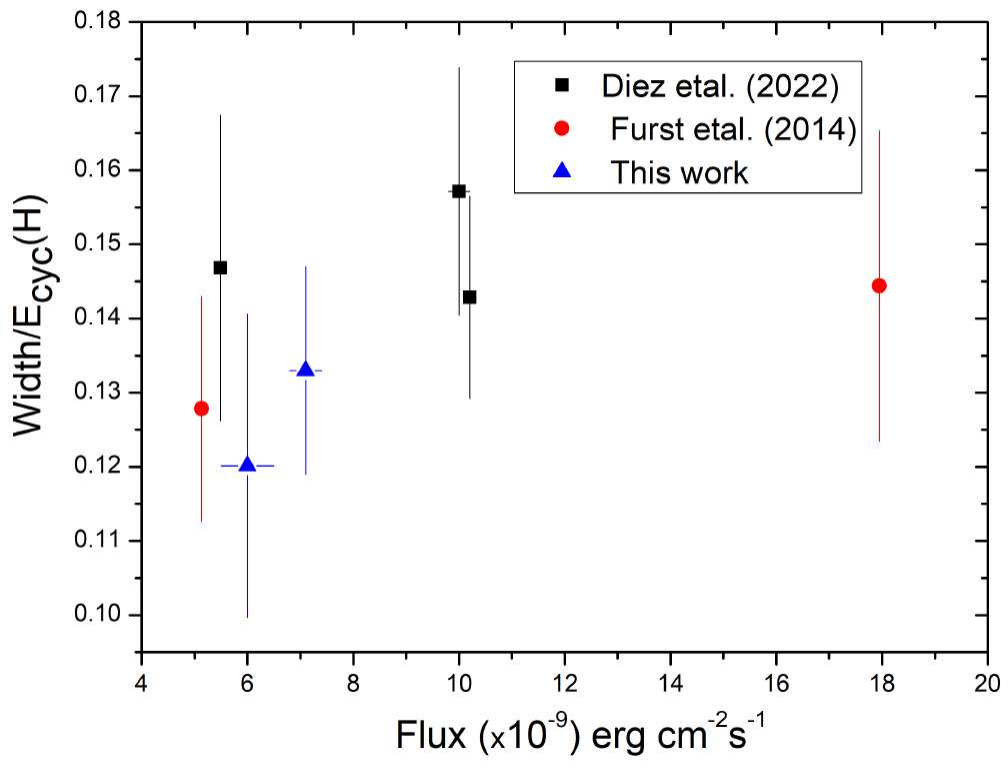}
\end{minipage}
\hspace{0.04\linewidth}
\begin{minipage}{0.30\textwidth}
\includegraphics[angle=0,scale=0.25]
{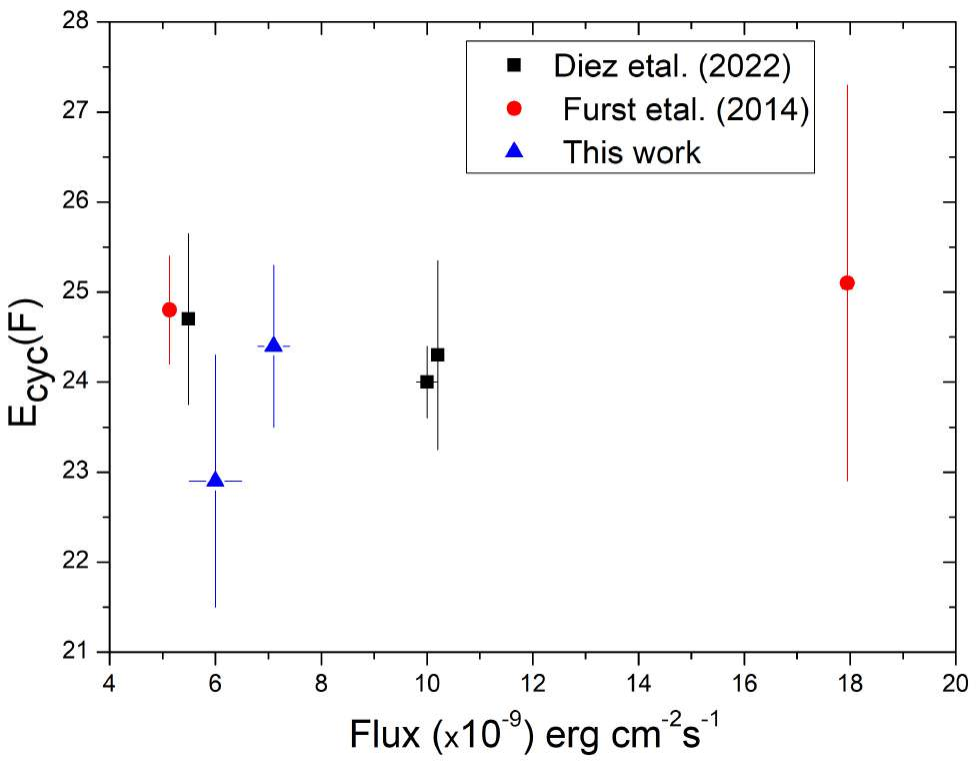}
\end{minipage}
\caption{Flux dependence of CRSF harmonic parameters:  line centroid $E_{\rm cyc,H}$,  relative width $\sigma/E_{\rm cyc,H}$, and  fundamental line centroid $E_{\rm cyc,F}$. Data from this work are compared with earlier results. Uncertainties are at the 1 $\sigma$ level.}
\label{6}
\end{figure*}

\begin{figure*}
\begin{minipage}{0.3\textwidth}
\includegraphics[angle=0,scale=0.28]{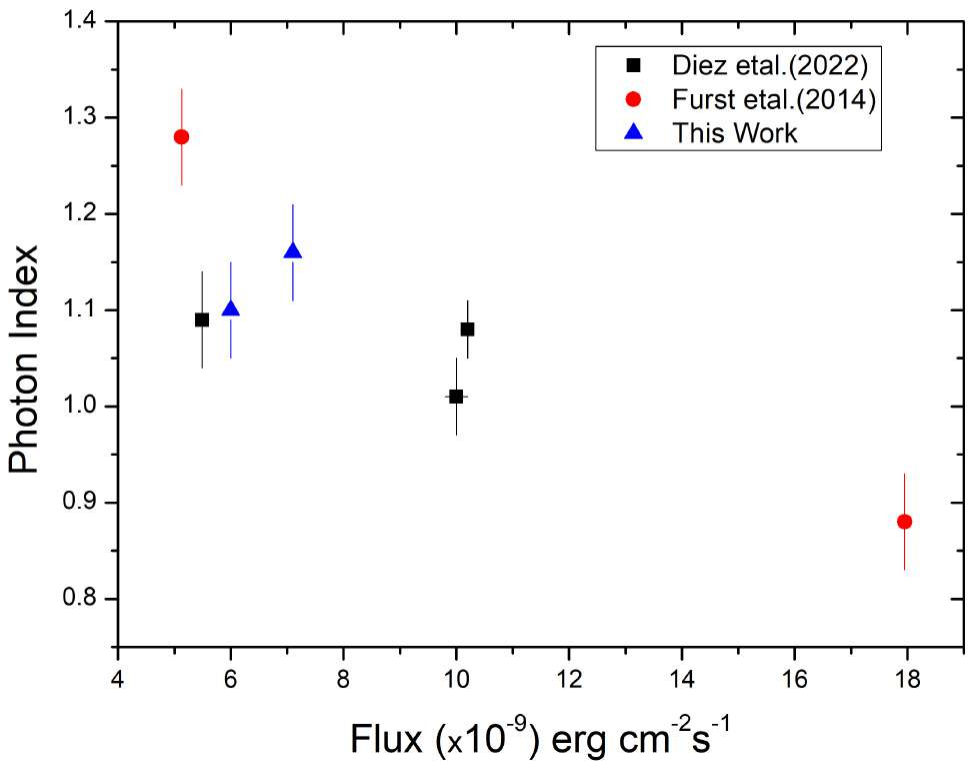}
\end{minipage}
\hspace{0.20\linewidth}
\begin{minipage}{0.3\textwidth}
\includegraphics[angle=0,scale=0.28]{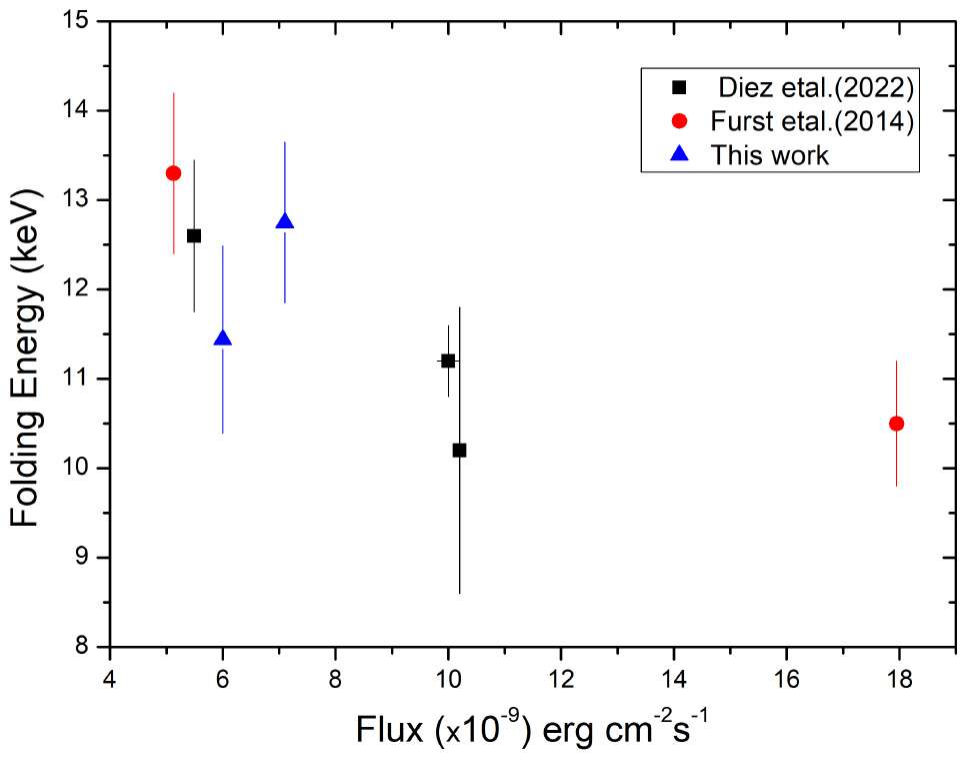}
\end{minipage}
\caption{Dependence of (left) photon index and (right) folding energy on flux. Errors  are at the 1 $\sigma$ level.}

\label{7}
\end{figure*}

\begin{table*}
\centering
\caption{Best-fit spectral parameters of Vela X-1 from \textit{NuSTAR} observations. 
Fluxes are quoted in the 3--79\,keV band. Errors represent 1$\sigma$ uncertainties. 
Subscripts F and H denote the fundamental and harmonic CRSFs, respectively.}
\label{tab:spec_params}
\begin{tabular}{lcc}
\hline
\textbf{Parameter} & \textbf{Obs I (2020-09-26)} & \textbf{Obs II (2020-09-29)} \\
\hline
Cross-normalization $C_{\rm FPMA}$ & 1.0 (fixed) & 1.0 (fixed) \\
Cross-normalization $C_{\rm FPMB}$ & $1.010 \pm 0.002$ & $0.997 \pm 0.003$ \\[3pt]

$N_{\rm H,1}$ ($10^{22}$\,cm$^{-2}$) & 0.371 (fixed) & 0.371 (fixed) \\
$N_{\rm H,2}$ ($10^{22}$\,cm$^{-2}$) & $17.8 \pm 0.7$ & $18.4 \pm 1.3$ \\
Partial covering fraction $f_{\rm pc}$ & $0.764 \pm 0.008$ & $0.80 \pm 0.05$ \\[3pt]

Photon index $\Gamma$ & $1.16\pm 0.05$ & $1.10 \pm 0.05$ \\
Cutoff energy $E_{\rm cut}$ (keV) & $16.9 \pm 4.7$ & $14.9 \pm 0.9$ \\
Folding energy $E_{\rm fold}$ (keV) & $12.75 \pm 0.90$ & $11.44 \pm 1.05$ \\[3pt]

Fe $K_\alpha$ centroid $E_{\rm Fe}$ (keV) & $6.343 \pm 0.062$ & $6.321 \pm 0.021$ \\
Fe line width $\sigma_{\rm Fe}$ (keV) & $0.110 \pm 0.001$ & $0.144\pm 0.004$ \\[3pt]

10 keV absorption $E_{\rm gabs}$ (keV) & $10.31 \pm 0.20$ & $9.88 \pm 1.05$ \\
$\sigma_{\rm gabs}$ (keV) & $1.8 \pm 0.3$ & $2.6\pm 1.1$ \\[3pt]

$E_{\rm cyc,F}$ (keV) & $24.4 \pm 0.9$ & $22.9 \pm 1.4$ \\
$\sigma_{\rm cyc,F}$ (keV) & $2.4 \pm 0.6$ & $2.1 \pm 0.5$ \\
Strength$_{\rm cyc,F}$ (dimensionless) & $0.30 \pm 0.10$ & $0.32 \pm 0.09$ \\[3pt]

$E_{\rm cyc,H}$ (keV) & $57.85 \pm 1.01$ & $54.11\pm 1.03$ \\
$\sigma_{\rm cyc,H}$ (keV) & $7.7 \pm 0.8$ & $6.5 \pm 1.1$ \\
Strength$_{\rm cyc,H}$ (dimensionless) & $22.4 \pm 4.6$ & $12.2 \pm 1.9$ \\[3pt]

Flux (3--79 keV) ($10^{-9}$\,erg\,cm$^{-2}$\,s$^{-1}$) & $7.1 \pm 0.3$ & $6.0 \pm 0.5$ \\[3pt]

$\chi^2$/d.o.f & 1173.9/1083 & 1157.6/1048 \\
Reduced $\chi^2_\nu$ & 1.08 & 1.10 \\
\hline
\end{tabular}
\end{table*}

\subsection{Evolution of CRSF and Continuum Parameters}

We examined the dependence of the cyclotron resonant scattering feature (CRSF) parameters on flux. In our analysis, no evidence of a correlation was found between the harmonic centroid energy and flux, as previously reported by \citet{furstt}. Similarly, we did not detect any correlation of the fundamental CRSF parameters with flux. A positive correlation between line energy and luminosity has already been questioned by \citet{Diez} based on orbit-resolved data, and our results support this conclusion.  

On the other hand, the photon index, which represents the spectral slope, shows a clear anti-correlation with flux, becoming harder as flux increases \citep{40, Klochkov}. The folding energy exhibits a similar trend of decreasing with flux. These results, combined with earlier findings in the literature, are shown in Figure \ref{6} for the CRSF parameters and in Figure \ref{7} for the photon index and the folding energy.  

\begin{figure*}

\begin{minipage}{0.4\textwidth}
\begin{center}
\includegraphics[angle=0,scale=0.3]{cych_r2.pdf}
\end{center}
\end{minipage}
\hspace{0.1\linewidth}
\begin{minipage}{0.4\textwidth}
\begin{center}
\includegraphics[angle=0,scale=0.335]{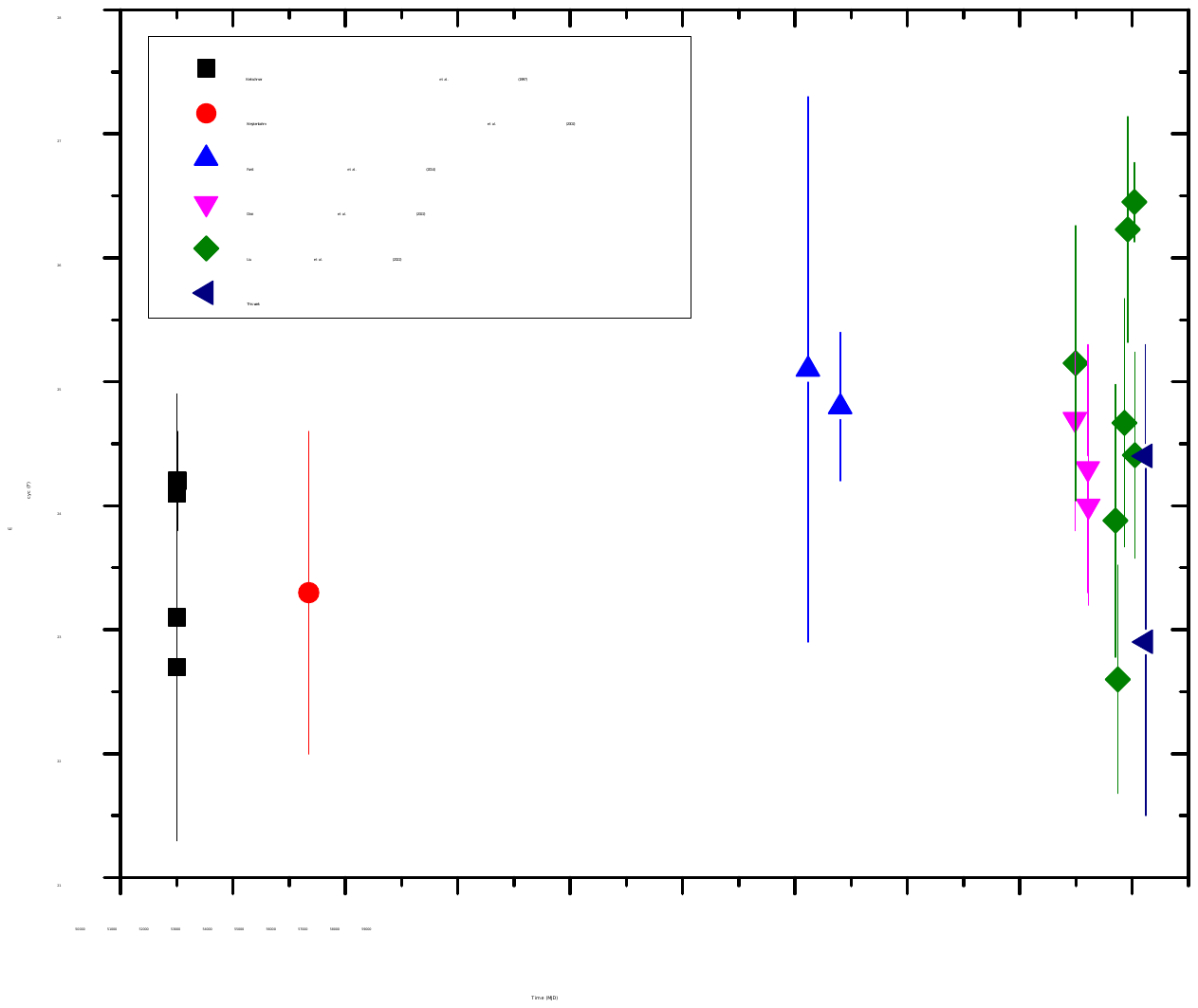}
\end{center}
\end{minipage}
\caption{Long-term evolution of CRSF energies in Vela X-1. Left: harmonic centroid energy $E_{\rm cyc,H}$, showing variability but no persistent decay. Right: fundamental centroid energy $E_{\rm cyc,F}$, with complex evolution from 2012--2020. Uncertainties are at the 2 $\sigma$ level.}

\label{8}
\end{figure*}

We also investigated the long-term evolution of the harmonic CRSF energy in the context of the \textit{Swift}/BAT monitoring results reported by \citet{La Parola} and \citet{Ji}. Earlier studies suggested that Vela X-1 may be one of the few sources showing long-term decay in the centroid energy of the harmonic line. Measurements with \textit{RXTE} up to 2016 hinted at such a trend. To test this, we included more recent \textit{NuSTAR} observations from 2019 and 2020. The temporal evolution of the fundamental line energy $E_{\rm cyc(F)}$ and the harmonic $E_{\rm cyc(H)}$ is presented in Figure \ref{8}.  

Our analysis shows that after 2016, the decaying trend of $E_{\rm cyc(H)}$ is no longer evident. For the 2019 \textit{NuSTAR} observations, $E_{\rm cyc(H)}$ was measured at 53.8 keV, 51.8 keV, and 56.0 keV. In 2020, the harmonic line energies were found at 57.9 keV and 54.2 keV. These results argue against a persistent long-term decay and instead suggest short-term variability, including a sudden rise in $E_{\rm cyc(H)}$ from 51.8 keV in 2019 to 57.9 keV in 2020.  

We also investigated the long-term evolution of the fundamental line $E_{\rm cyc(F)}$. According to \citet{Kretschmar}, the fundamental line was measured in the range 22.7--24.2 keV in 1997. Subsequent measurements around 2000 found $E_{\rm cyc(F)} = 23.3\pm1.3$ keV \citep{Kreykenbohmm}, with no clear evidence of a decaying trend. From 2012 to 2020, however, \textit{NuSTAR} observations reveal a more complex pattern of variability in $E_{\rm cyc(F)}$, as illustrated in Figure \ref{8}. The current data are insufficient to establish a long-term secular trend for the fundamental line, largely due to limited detections.

\subsection{Continuum Evolution and Time-Resolved Spectroscopy}

\begin{figure}

\includegraphics[angle=0,scale=0.34]{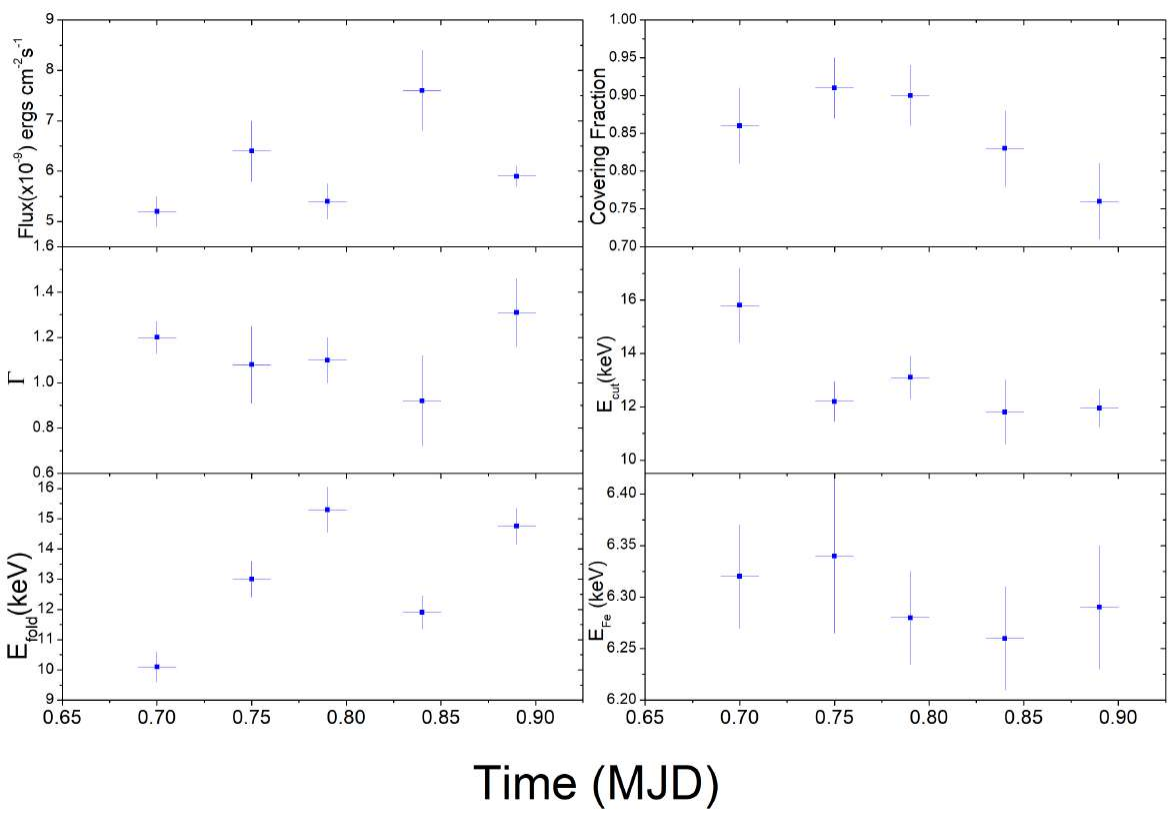}

\caption{Variation of spectral parameters: Flux, Partial covering fraction, Photon Index ($\Gamma$), Cut-off energy ($ E_{cut}$), folding energy ($ E_{fold}$) and iron line energy with time in the range (59121.69 - 59121.92) MJD labelled as 0.69 - 0.92 along the X-axis for NuSTAR observations of Vela X-1. Horizontal error bars denote the integration time of the spectra. Uncertainties are at the 1 $\sigma$ level.}
\label{9}
\end{figure}

We investigated the temporal evolution of the continuum and spectral parameters by dividing the \textit{NuSTAR} light curves into several segments. The light curve was divided into a series of non-overlapping time segments spanning the full observation. Segment boundaries were selected based on the natural variations in  the light curve, while keeping the exposure in each interval broadly comparable. For each segment, spectra were extracted using good time interval (GTI) files, and the same continuum model adopted for the phase-averaged analysis was applied. The evolution of the spectral parameters for observation II is shown in Figure \ref{9}.  

The source flux in the 3--79 keV range varies between  (5.20 - 7.54) $\times 10^{-9}$ erg cm$^{-2}$ s$^{-1}$, indicating significant variability between segments. The partial covering fraction of the \texttt{TBPCF} component is found to fluctuate in the range 0.75--0.99, consistent with the presence of variable absorption from the clumpy stellar wind. The cutoff and folding energies also exhibit noticeable changes: $E_{\rm cut}$ ranges from 12.0--15.7 keV, while $E_{\rm fold}$ varies between 11.9--15.5 keV. The centroid of the Fe $K_\alpha$ fluorescence line shows modest variations between 6.28 and 6.35\,keV, consistent within statistical uncertainties (Figure \ref{9}).  \textbf{While residual NuSTAR calibration uncertainties in the Fe-K band may contribute to the observed variations, the line is consistent with near-neutral iron within the uncertainties.}

\begin{figure}
\begin{center}
\includegraphics[angle=0,scale=0.34]{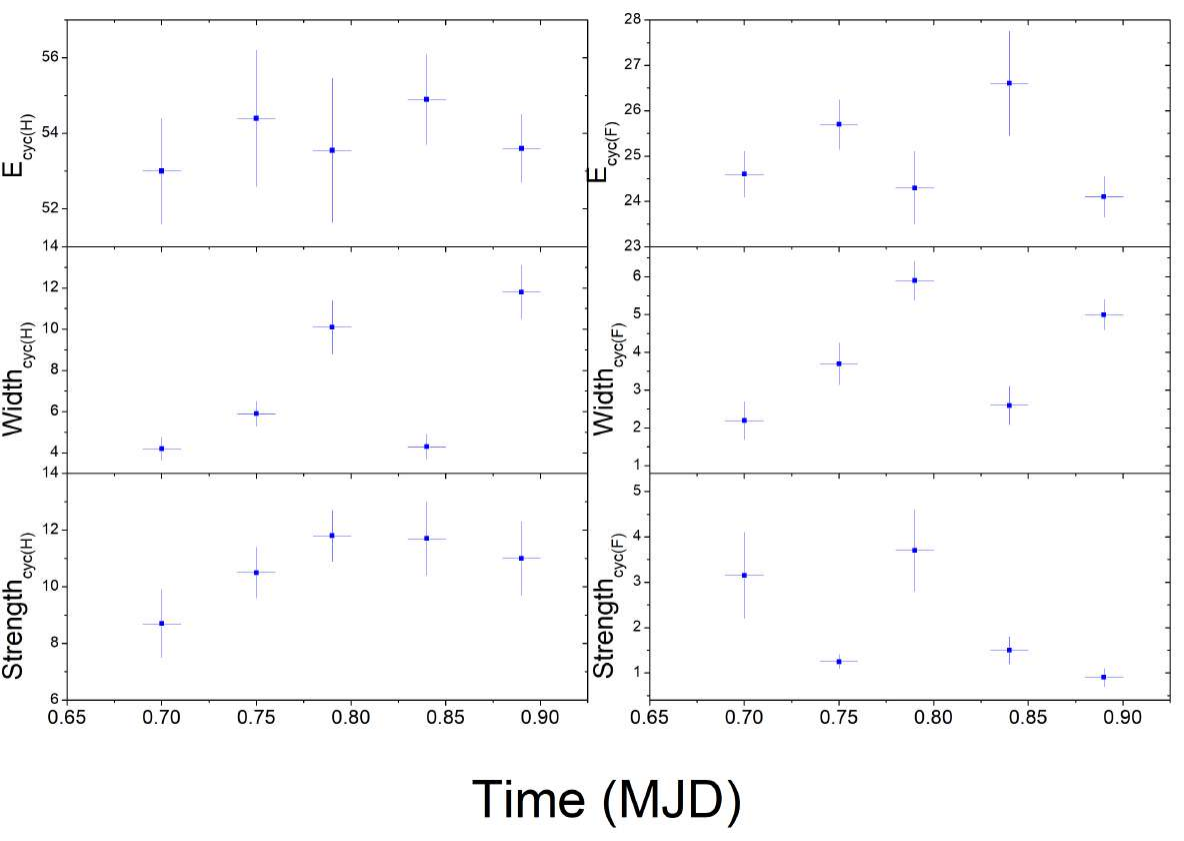}
\caption{Variation of cyclotron line parameters (energy, width and strength in keV) with time in the range (59121.69 - 59121.92) MJD labelled as 0.69 - 0.92 along the x-axis.) MJD. Horizontal error bars denote the integration time of the spectra. Uncertainties  are at the 1 $\sigma$ level.}
\label{10}
\end{center}
\end{figure}

Cyclotron features are prominently detected in the spectra of all segments, with the line parameters exhibiting measurable evolution (Figure \ref{10}). Most continuum parameters show only moderate variability; however, the centroid energy of the fundamental CRSF deviates by up to $\sim 5\%$ and that of the harmonic by $\sim 2\%$ relative to their phase-averaged values. The fundamental line reaches a maximum of  26.6 keV in segment 4 and a minimum of  24.4 keV in segment 3. The harmonic line attains a maximum of  54.8 keV in segment 4 and a minimum of  53.0 keV in segment 1.  

\begin{figure}
\begin{center}
\includegraphics[angle=0,scale=0.25]{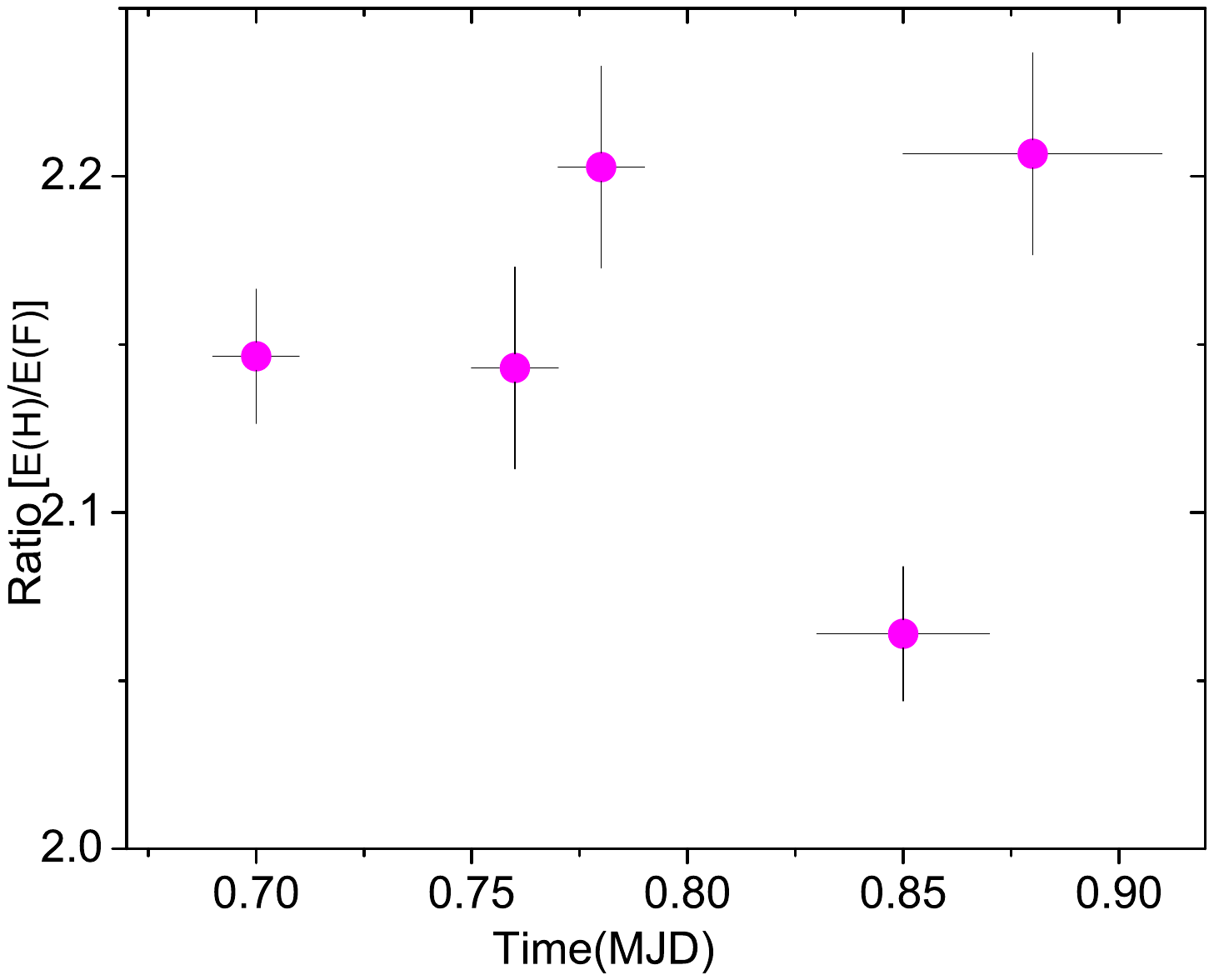}
\caption{Ratio of harmonic to fundamental CRSF energies, $E_{\rm cyc,H}/E_{\rm cyc,F}$, as a function of time during Observation~II. The values consistently exceed the canonical value of two, indicating anharmonic spacing.}
\label{11}
\end{center}
\end{figure}

As illustrated in Figure \ref{11}, the ratio of harmonic to fundamental centroid energies, $E_{\rm cyc(H)}/E_{\rm cyc(F)}$, remains consistently anharmonic (greater than the classical value of 2). This suggests that the line-forming regions for the fundamental and harmonic features may be located at different heights within the accretion column.  


\section{Discussion}

We conducted a detailed temporal and spectral analysis of two \textit{NuSTAR} observations of Vela X-1, complemented by four archival datasets covering nearly a decade. The recent data from 2020 provide new perspectives on pulse morphology, spectral variability, and the behaviour of cyclotron lines in this system. 

\subsection{Pulse Profiles \& Accretion Geometry}
The observed transformation of pulse profiles—from irregular structures in the soft band to a well-defined double-peaked structure at higher energies—indicates changes in the emission geometry with photon energy.  This is expected in wind-fed accreting systems at moderate luminosity. The unequal peak amplitudes and phase separation suggest that either the magnetic poles are not strictly antipodal or the inflowing material is channeled in an asymmetric manner. The pulse fraction (PF) increases with energy but decreases with flux, which is consistent with a scenario where the unpulsed thermal component grows stronger during enhanced accretion rates, diluting the modulated emission from the accretion column.

\subsection{Cyclotron Features \& Magnetic Field}
Both the fundamental and harmonic cyclotron resonant scattering features (CRSFs) are observed in our analysis. Although the fundamental line is weak, it is likely suppressed by photon spawning \citep{Schwarm} and blending with the stronger harmonic. From the measured centroid energies, the surface magnetic field of the neutron star is estimated to be $B \approx (2.7-2.8)\times10^{12}$\,G, in agreement with previous estimates. Unlike earlier reports based on different observatories, we do not find evidence for a secular decay in the harmonic CRSF energy. A noteworthy result is that the ratio $E_{\rm cyc(H)}/E_{\rm cyc(F)}$ is consistently greater than two, suggesting that the two features originate at different altitudes within the accretion column. Such anharmonicity provides important constraints on the local magnetic field configuration and the extent of the line-forming region. This trend is also verified using time-resolved spectroscopy. 

\subsection{Long-term Behaviour of CRSFs}
Earlier studies hinted at a gradual decline in the harmonic line energy, similar to the secular evolution seen in Her~X-1. Our updated analysis does not confirm this trend. Instead, we find that the harmonic energy undergoes short-term fluctuations, including a marked increase between the 2019 and 2020 observations. The fundamental line shows a complex evolution without a simple monotonic trend. These patterns suggest that local structural changes in the accretion column or mound---such as variations in height, density, or magnetic topology---play a dominant role, rather than a global decay of the neutron star's dipole field. Differences between \textit{NuSTAR} and \textit{Insight-HXMT} measurements may arise from calibration uncertainties or instrument-dependent sensitivities, and require further cross-mission studies. 

\subsection{Continuum Evolution and Sub-critical Accretion}
The spectral slope becomes harder with increasing flux, as revealed by the anti-correlation of both the photon index and folding energy with luminosity. This is a characteristic of accretion in the sub-critical regime, where higher mass inflow leads to more efficient Comptonization in the column plasma. The measured luminosities, of order $10^{36}$\,erg\,s$^{-1}$, are well below the theoretical critical value, supporting the interpretation that Vela X-1 remains firmly in the sub-critical accretion state. Time-resolved spectroscopy further shows that CRSF energies vary on short timescales, reinforcing the idea that the line-forming region is sensitive to changes in local accretion conditions.

 
\subsection{Time-Resolved Spectroscopy and Anharmonicity of Cyclotron Features}

To investigate the spectral evolution of Vela X-1, we performed time-resolved spectroscopy. The continuum parameters exhibit noticeable variability across different segments, and the cyclotron features also evolve with time.  

In a simple picture, the electron energy levels in a strong magnetic field are quantized into Landau levels, and the centroid energy of the first harmonic is expected to be exactly twice that of the fundamental line. However, our results show that the harmonic energy exceeds twice the fundamental value, indicating anharmonic spacing. Similar deviations have been reported in a few other X-ray pulsars. Small departures from the canonical ratio of 2 can be explained by relativistic corrections to photon-electron scattering \citep{Meszaros,Schonherr}. However, relativistic effects alone are insufficient to explain the full extent of the observed anharmonicity.  

Several additional mechanisms may contribute. If the fundamental and harmonic lines form at different heights in the accretion column, variations in optical depth may shift their apparent centroid energies. A decrease in the magnetic field strength with altitude within the line-forming region can also increase the observed line ratio \citep{Nishimura,Schonherr}. In such cases, the superposition of fundamental lines originating at multiple heights can shift the average energy downward, while the harmonic remains relatively stable. Furthermore, deviations from a pure dipole field geometry may distort the field configuration and modify the line energies, producing anharmonic ratios.  

For wind-accreting pulsars such as Vela X-1, the critical luminosity $L_{\rm c}$, which separates the sub-critical and super-critical accretion regimes, is expected to be higher than for disk-accreting systems \citep{Mushtukov}. The critical luminosity can be approximated as  

\begin{equation}
L_{\rm c} \approx 1.5 \times 10^{37} \, B_{12}^{16/15} \; \mathrm{erg \; s^{-1}},
\end{equation}

where $B_{12}$ is the surface magnetic field in units of $10^{12}$ G \citep{Beckerr}. The luminosity values derived from our \textit{NuSTAR} observations are well below $L_{\rm c}$, consistent with Vela X-1 accreting in the sub-critical regime.  

\section{Summary and Conclusion}

Vela X-1 has been extensively observed with various X-ray observatories, providing a foundation for studies of its timing and spectral properties, including the behaviour of cyclotron resonant scattering features (CRSFs). In this work, we analyzed two relatively new \textit{NuSTAR} observations from 2020 along with four earlier datasets spanning 2012--2020.  

Our main findings are summarized as follows:  

\begin{itemize}
    \item We do not confirm the previously reported correlation between the centroid energy of the harmonic line and source luminosity, consistent with the relatively recent studies that question the robustness of this trend \citep[e.g.][]{Diez}.  
    \item The long-term decrease in the harmonic CRSF energy suggested in earlier studies is not supported by these observations. Instead, the harmonic line shows short-term variability without a clear secular trend.  
    \item The fundamental CRSF energy exhibits a complex and irregular evolution during 2012--2020, with no clear long-term monotonic behaviour.  
    \item No significant correlation is found between CRSF parameters and source flux.  
    \item The photon index and folding energy display a decreasing trend with increasing flux, consistent with spectral hardening due to changes in the Comptonizing plasma.    
    \item The average accretion rate during the 2020 \textit{NuSTAR} observations is estimated to be of order $\sim 10^{16}$ g s$^{-1}$.    
    \item The pulse fraction increases with photon energy but decreases with flux, consistent with wind-fed accretion models.  
    \item The ratio of harmonic to fundamental centroid energies departs from the canonical value of 2, suggesting that the line-forming regions are located at different heights in the accretion column.  
\end{itemize}  

Given the importance of Vela X-1 as a persistent wind-accreting system, continued monitoring is strongly encouraged. Future observations with current and next-generation X-ray missions, combined with improvements in theoretical modeling, will be essential for clarifying the long-term behaviour of its cyclotron features and the physics of sub-critical accretion in strongly magnetized neutron stars.

\section{Acknowledgements}
This research work is done by using publicly available data provided by NASA HEASARC data archive. The NuSTAR data is provided by NASA High Energy Astrophysics Science Archive Research Center (HEASARC), Goddard Space Flight Center. We are grateful to IUCAA Centre for Astronomy Research and Development (ICARD), Department of Physics, NBU, for the research facilities. MT acknowledges CSIR, India for the research grant 161-411-3508/2K23/1. We express our gratitude to the anonymous reviewer for the valuable comments and suggestions that have substantially improved the manuscript. 
 
\section{Data availability}
 
The observational data used for performing this research work can be accessed from the HEASARC data archive and is publicly available for analysis.








\end{document}